# TEMPERATURE DEPENDENT OPTICAL RESPONSE OF High-$T_c$ YBa$_2$Cu$_3$O$_{7-\delta}$ (YBCO) THIN FILMS

Vivek Khichar[1,2], V. A. Chirayath[1,2], Jonathan Asaadi[1,2], Nader Hozhabri[3,a], Benjamin J. P. Jones[1,2], Ali R. Koymen[1], Iakovos Tzoka[1], Pratyanik Sau[1]

[1]Department of Physics, University of Texas at Arlington, Arlington, Texas 76019, USA

[2]Center for Advanced Detector Technology, University of Texas at Arlington, Arlington, Texas 76019, USA

[3]Nanotechnology Research Center, Shimadzu Institute, University of Texas at Arlington, Arlington, Texas 76019, USA

[a]Author to whom correspondence should be addressed: nh@uta.edu

# ABSTRACT

We report on the temperature-dependent optical response of thin films of $YBa_2Cu_3O_{7-\delta}$ (YBCO) in the visible spectral range under cryogenic conditions. Specifically, we observe an increase in transmittance near the superconducting transition temperature ($T_c$), which saturates within a few kelvins below $T_c$. The increase in transmittance is accompanied by a corresponding decrease in reflectance as the temperature drops below $T_c$, and both quantities track the superconducting phase transition. Changes in transmittance are found to be wavelength dependent, with the maximum variation occurring at 633 nm and minimal at 450 nm. These observations establish a correlation between the variation in optical response and the superconducting phase transition, even in the visible regime. The results of our experiment highlight the potential for using non-contact optical measurements to determine $T_c$. The effect can be explained using the two-fluid model, which can account for the observed temperature and wavelength dependence of the transmittance of the superconducting thin films.

# I. INTRODUCTION

Yttrium barium copper oxide (YBa2Cu3O$_{7-\delta}$, or YBCO) is a pioneering high-temperature superconductor (HTS), distinguished by its $T_c$ above 90 K [1]. This temperature, which exceeded the boiling point of liquid nitrogen (77 K), heralded the research into high-$T_c$ superconductivity (HTSC). The interaction between the alternating CuO$_2$ planes with the quasi-one-dimensional Cu-O chains in the orthorhombic perovskite structure of YBCO has been shown to be important in enabling superconductivity in YBCO [2]. The oxygen content within the YBCO structure also plays a crucial role in determining its superconducting properties, transitioning from an insulating tetragonal phase when the oxygen content is low to a superconducting orthorhombic phase at optimal doping levels [3].

Beyond its superconducting capabilities, YBCO thin films have garnered attention for their optical properties, including strong anisotropic behavior and, observation of surface plasmons in the infrared region [4-9]. The optical behavior, characterized by properties such as reflectance, transmittance, and absorption, provides information on the electronic band structure and quasi-particle excitations in the material [10, 11]. Recent advances in thin-film deposition techniques have allowed precise control over morphology and stoichiometry, facilitating detailed exploration of these optical properties under varied conditions [12, 13].

Fishman et al. [14] showed that the differential optical reflectance of YBCO measured at 780 nm has a sharp peak at $T_c$ with a width of less than 1 K providing a way to optically measure inhomogeneities in $T_c$ and the dynamics of phase transition across the superconducting thin film. Unlike traditional electrical methods, such optical characterization offers a non-contact diagnostic approach, potentially enabling the development of novel superconducting based optical sensors and detectors based on light-matter interactions.

Here we present the optical properties of high-$T_c$ YBCO thin films under visible light (450 nm to 633 nm) measured as a function of sample temperature. Our experimental results demonstrate clear changes in the optical properties of YBCO transitioning to the superconducting state. These findings are further supported by an empirical two-fluid model, which qualitatively describes the optical response of the material when the thin film transitions from a normal state to a superconducting state. The model provides a plausible explanation for the observed changes and establishes a foundation for using optical techniques for measuring $T_c$.

The manuscript is constructed as follows: section II describes the experimental setup used to measure the optical and electrical characteristics of YBCO thin films as a function of sample temperature, and provides details of the optical cryogenic system, sample characterization, and correction factors incorporated in optical measurements. Section III presents the experimental data, highlighting the optical transitions associated with the superconducting phase transition and the reproducibility of these effects. Section IV discusses the predictions of the two-fluid model and their alignment with experimental observations, followed by section V which concludes the work by discussing the implications of our findings and its potential applications as an optical readout scheme in superconducting devices.

## II. EXPERIMENTAL DETAILS

### A. Cryogenic optical spectroscopy and electrical measurements

Figure 1 shows the schematic of our closed-cycle optical cryogenic apparatus capable of simultaneously measuring the optical response of the sample (reflectance and transmittance measurement) and its electrical properties (resistivity measurements) as a function of sample temperature.

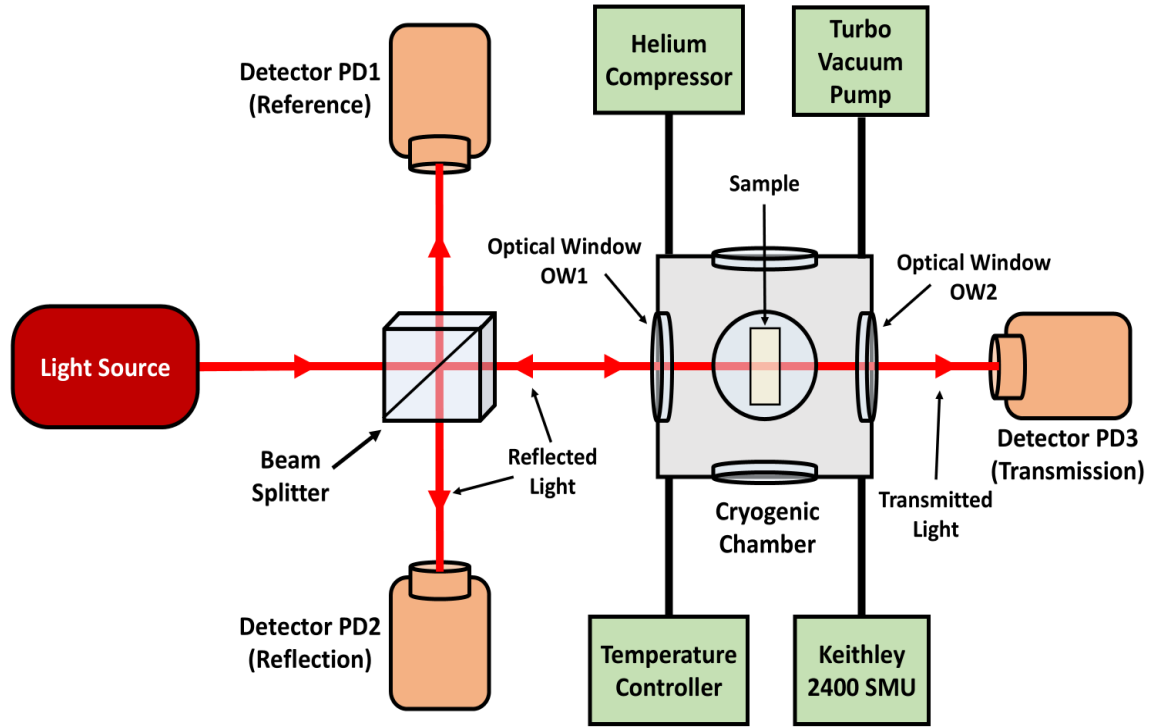

Figure 1. Block diagram of the experimental set up (as a top view) for simultaneous measurements of the reflected and transmitted light from the sample under cryogenic conditions. Light beam from a light source is split as Reference and Incoming beam using a beam splitter. The reference beam is detected using photodetector PD1. The incoming beam passes through an Optical Window OW1 of the cryogenic chamber and is incident upon the sample. Light reflected back from the sample and optical windows are collected in photodetector PD2. The transmitted light is detected at PD3. A Keithley SourceMeter Unit (SMU) is coupled to the in-built Four-point Probe assembly and the resistivity of the sample can be measured.

In our custom-designed optical cryogenic system (from ColdEdge Technologies), the sample can be mounted horizontally (i.e. the sample lies flat on the copper cold finger) or vertically (the sample is mounted on one of its edges on a custom-built sample holder on the copper cold finger). Closed-cycle cryogenic equipment allows us to cool the sample to ~ 10 K. The cryostat also uses an integrated heating element that allows us to heat the sample up to 200 °C. The cryostat employs the APD Cryogenics HC-4 MK2 helium compressor for the circulation of ultra-high purity helium. The temperature of the system is controlled using the LakeShore Model 335 cryogenic temperature controller. The temperature is monitored using a Si-diode (DT-670) sensor located on the copper cold-finger head in the close vicinity of the mounted sample. The optical cryo-chamber is evacuated to a base pressure of $\leq 10^{-7}$ Torr using a turbo pump before cooling the system.

For optical studies, the sample is mounted vertically on the cold finger. As shown in figure 1, light beam from the light source is split by a beam splitter. One portion is directed to photodetector PD1 and recorded as the reference signal. The other portion passes through optical window OW1 and reaches the sample as the incident beam. The reflected light travels back through OW1 to the beam splitter and is then directed to photodetector PD2, while the transmitted light exits through a second optical window OW2 and is measured by photodetector PD3. These signals correspond to the reflectance and transmittance, respectively. All optical windows of the system are made of 1.25 inch diameter and 0.25 inch thick of quartz. All photodetectors used in the system are PASCO Scientific model CI-6504A Light Sensor which are Si-PIN photodiodes with spectral response range from 320 nm to 1100 nm. All reflectance and transmittance measurements are carried out simultaneously and remotely using the PASCO Scientific ScienceWorkshop 750 model interface connected to a computer using PASCO Scientific DataStudio software.

For electrical studies, the sample is mounted horizontally on the cold finger and resistivity is measured by a Keithley model 2400 SMU and a custom designed built-in Four-point Probe assembly mounted on the cold finger. This assembly uses 0.508 mm radius tungsten tips with 1.016 mm spacing between the tips. The resistivity measurements were performed using the SMU in-built Four-Point Probe resistivity measurement functionality.

The experimental work presented here weas performed in three steps.

- Initially (Section IIIA), electrical measurements of the YBCO films were performed, and the resistivity of the film as a function of sample temperature was measured.
- In second part (Section IIIB), optical measurements of the YBCO films were performed using a mixed polarized He-Ne laser light of wavelength $\lambda \sim 633\ nm$ as a light source for simultaneous measurements of reflectance and transmittance as a function of sample temperature at near normal incidence of light upon sample surface.
- Finally (Section IIIC), a mixed-polarized variable wavelength laser (SuperK EXTREME from NKT Photonics) was used as a light source for measuring the transmittance of the YBCO films at 450 nm, 500 nm, 530 nm, 580 nm, and 630 nm at normal incidence and as a function of sample temperature.

### B. Thin film detail and characterization

The high-$T_c$ superconducting epitaxial thin films used in these experiments were commercially obtained and grown with a <100> orientation on single-crystal SrTiO3 (STO) substrates, also oriented along <100> [15]. These films were deposited on double-side polished STO substrates with dimensions of 10 mm × 10 mm × 0.5 mm at a thickness of 100 nm. According to the vendor, the $T_c$ of the films is $90\ K \pm 5\%$ ($86\ K - 90\ K$).

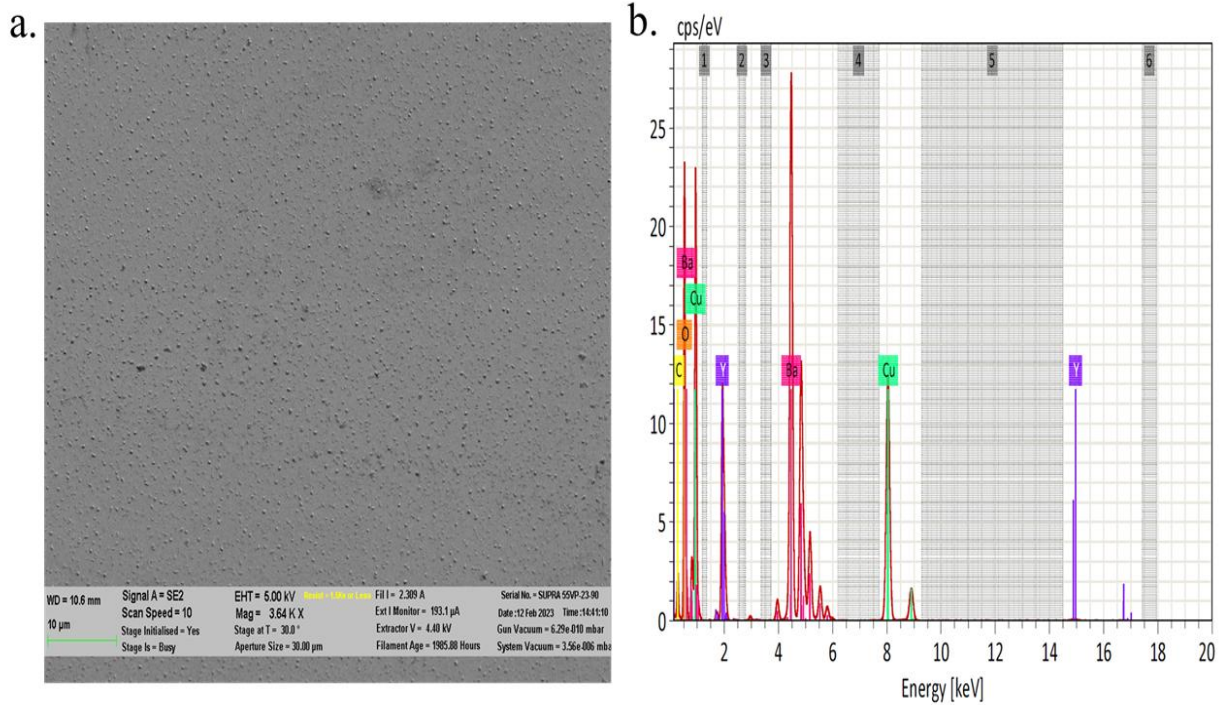

Figure 2. SEM and EDS scans of 100 nm YBCO thin film. a. SEM image shows a relatively smooth surface at 10 µm scale. The images taken after the cold cycle runs did not show any cracks due to differential thermal expansion between the film and the substrate. b. EDS data obtained from the same region as shown in the SEM image with the characteristic x-ray peaks from various elements (Y, B, Cu and O) marked. EDS data confirmed that the metal ions occur in the 1:2:3 composition in the YBCO thin film. The oxygen concentration in as received sample was 6.84 which reduced to 5.83 after keeping the sample in vacuum for multiple days during the measurements.

We used Scanning Electron Microscopy (SEM) and Energy-dispersive X-ray spectroscopy (EDS) (shown in figure 2) to examine surface topography and atomic composition of the YBCO thin films before and after cryogenic runs. EDS of the as received samples confirmed a stoichiometry consistent with $YBa_2Cu_3O_{6.84}$. However, after the sample went through multiple high-vacuum cryogenic cycles the stoichiometry, as measured by EDS, changed to $YBa_2Cu_3O_{5.83}$. The reduction in oxygen content in high-vacuum environment also resulted in a reduction of $T_c$. To recover the oxygen content and to increase the $T_c$ of YBCO thin films, the sample was annealed at 500 °C for 12 Hours in $O_2$ ambient as per the vendor's instructions. The annealing process was done in a Linberg/Blue HTF55342-4 1200 °C digital Horizontal Oxidation furnace in the *class-100* clean room at the UTA Nanotechnology Research Center (NRC) facility.

### C. Correction factors in optical measurements

The light incident on, reflected from, and transmitted through the sample passes through optical windows (OW1 and OW2) and a beam splitter before reaching the corresponding photodetectors, as shown in figure 1. To determine the sample's reflectance and transmittance, we applied a set of intensity correction factors illustrated in figure 3. The transmittance of the quartz optical windows, $T_{ow}$, was calculated for normal incidence using the refractive index of quartz at the laser wavelengths. The beam splitter's transmission and reflection fractions — 0.19 (transmitted toward the sample), 0.30 (reflected toward PD1), and 0.28 (reflected toward PD2) — were measured independently using a calibrated power meter. The optical power incident on the sample inside the cryogenic chamber is given by $I = 0.19 T_{ow} I_0$, where $I_0$ is the laser output power in milliwatts. The reflection from the sample surface is then calculated as $R = \frac{R'}{0.30 T_{ow}}$, where $R'$ is the measured signal at PD2. Similarly, the transmission is obtained from $T = \frac{T'}{T_{ow}}$, with $T'$ being the signal measured at PD3. The final reflectance and transmittance were calculated as $\frac{R}{I}$ and $\frac{T}{I}$, respectively. These corrections account for all losses and splitting effects prior to detection, ensuring that the extracted values of reflectance and transmittance represent the optical response of the sample itself.

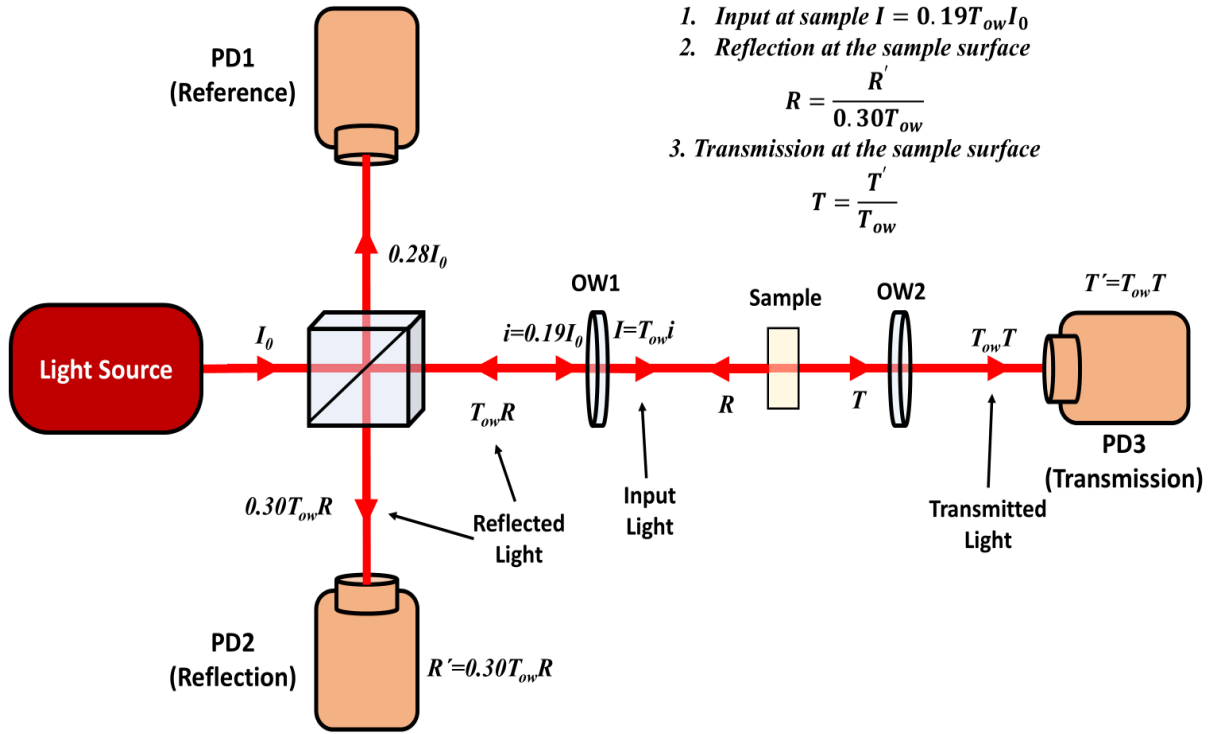

Figure 3. Schematic representation of all correction factors used in optical measurements. We note that the correction factors from the beam splitter is specifically for wavelength of λ ~ 633 nm.

## III. EXPERIMENTAL RESULTS

### A. Determination of $T_c$ using resistivity measurements

Figure 4 shows the measured normalized resistivity of the YBCO film as a function of sample temperature. The resistivity is normalized to the value at room temperature. The $T_c$ of the sample (obtained from an error weighted least squared fit with a logistic function which gave the mid-point of the sigmoid), is 87 ± 3 K, which is consistent with the vendor's specification of the $T_c$ and the value of 86.6 K obtained from two-fluid model.

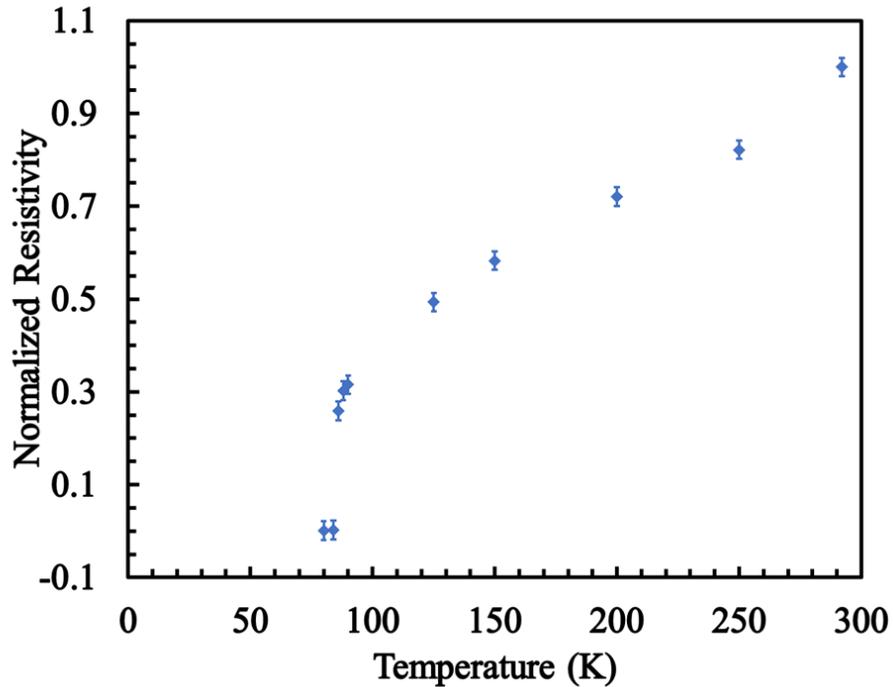

Figure 4. Normalized resistivity measurements of the YBCO thin film as a function of sample temperature. At room temperature the resistivity of the sample is measured to be 14.8 ± 0.002 µΩ · m. Note: Error bars are enhanced by 100 fold for better visibility.

### B. Single wavelength optical response

The optical response of the sample — specifically, its reflectance and transmittance — was measured as a function of sample temperature at a single wavelength of $\lambda \sim 633\ nm$, using near-normal incidence (≤ 1°). This small angle was chosen to help distinguish the light reflected from the sample surface from other reflections originating at the beam splitter and the front and back surfaces of the optical window (OW1).

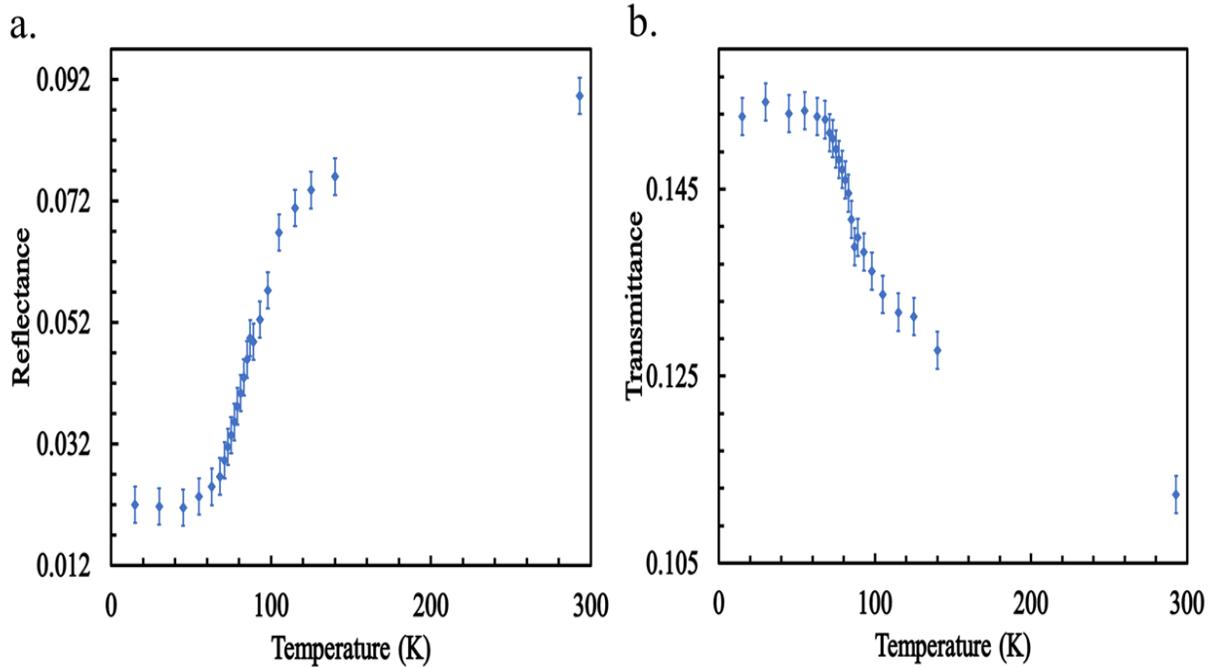

Figure 5. a. Reflectance and b. transmittance of light from YBCO thin film as a function of sample temperature at near normal incident angle for laser wavelength of $\lambda \sim 633\ nm$. Note: Error bars are enhanced for better visibility by a. three fold for reflectance and, b. two fold for transmittance.

Figure 5 shows the reflectance and transmittance of light from the YBCO thin film as a function of sample temperature. The superconducting transition is evident in both reflectance and transmittance data with the onset occurring around 100 K and reaching a saturation by 70 K. Compared to the room temperature value the reflectance drops by up to 76 % and transmittance increases by as much as 37 % during the transition from normal to superconducting state.

To confirm that the observed changes in reflectance and transmittance are intrinsic to the YBCO thin film and not due to temperature-dependent optical properties of the substrate, the refractive index of STO was measured as a function of sample temperature at $\lambda \sim 633\ nm$. A bare, double-side-polished single-crystal STO <100> was mounted vertically in the cryogenic chamber, and transmittance measurements were performed at normal incidence. Figure 6 shows the refractive index (n) of STO as a function of sample temperature, calculated from the transmittance data. The refractive index of STO exhibits no significant temperature dependence. Therefore, we conclude that the changes observed in figure 5 are attributable to variations in the optical properties of the YBCO thin film.

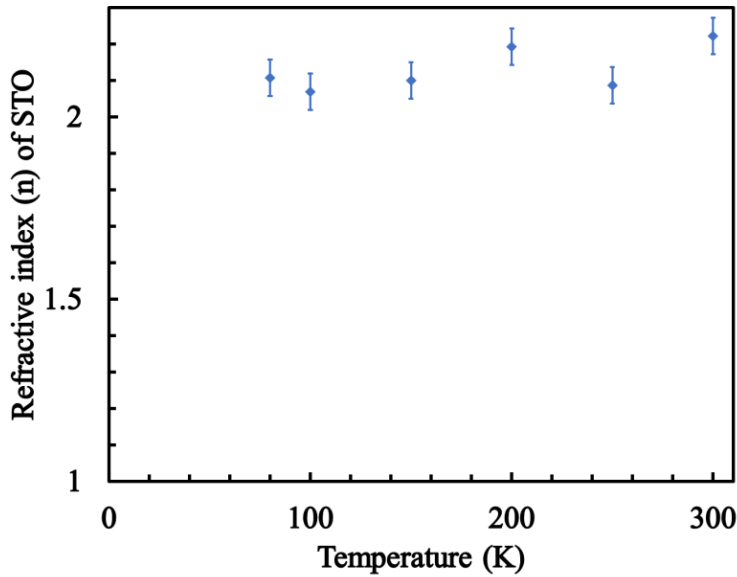

Figure 6. Refractive index (n) of STO sample as a function of sample temperature for wavelength of $\lambda \sim 633\ nm$, at normal incidence.

The reflectance and transmittance data were fitted using a logistic function to extract the midpoint of the sigmoid, corresponding to the optical determination of the $T_c$, as shown in figure 7. A weighted least-squares fit was performed, accounting for the experimental uncertainties. The extracted midpoint from both fitted curves is 82 ± 1 K, which is in close agreement with the $T_c$ obtained from electrical resistivity measurements, 87 ± 3 K, as shown in figure 4.

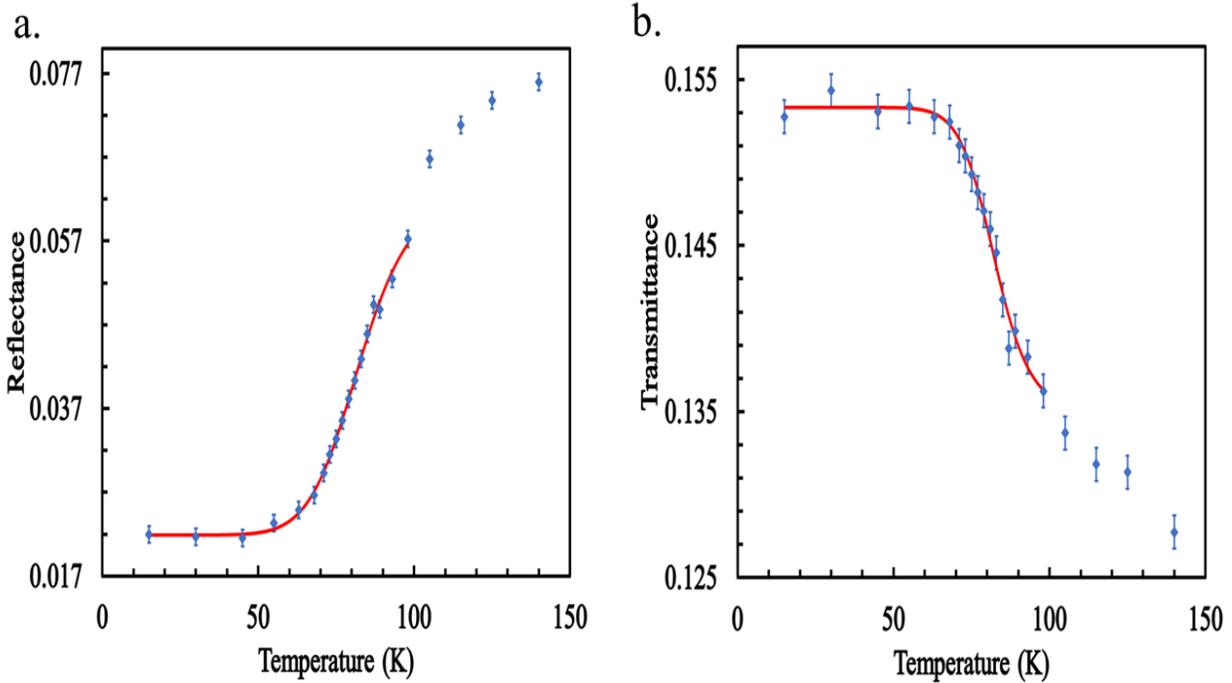

Figure 7. Determination of the critical transition temperature of YBCO thin film sample using a. reflectance data and b. transmittance data. Blue circles with error bars show reflectance and transmittance data with solid red line showing fitted Logistic curve.

As outlined previously, reflectance measurements require the incident beam to be directed at near-normal incidence to minimize multiple reflections originating in optical components other than the sample. However, this geometry can introduce polarization-dependent effects as well as make calculations of transmittance complex. To avoid such complications, subsequent experiments (described in Section IIIC) focused solely on transmittance measurements of the YBCO thin film as a function of sample temperature. In this configuration, the incident beam was exactly normal to the sample surface, and the absolute transmittance was calculated from the measured intensity at PD3 using a single correction factor, $T_{ow}$, accounting for the optical window. This factor depends only on the wavelength-dependent refractive index of the window material.

## C. Optical response at various wavelengths

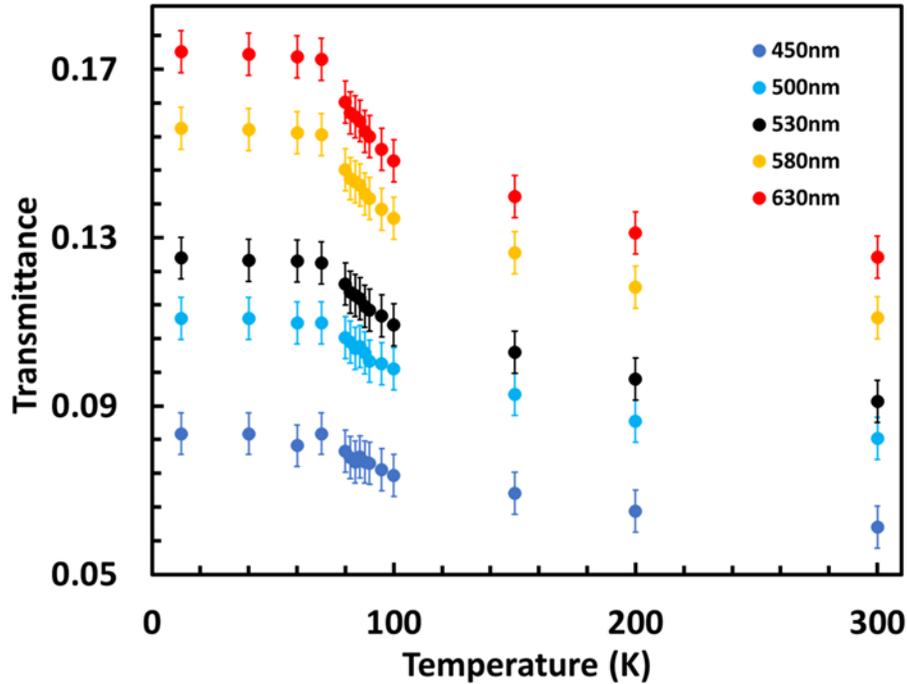

Figure 8. Transmittance of light from YBCO thin film sample as a function of sample temperature at various wavelengths. A variable wavelength laser, at normal incidence, was used as the light source. Note: Error bars are enhanced by five fold for better visibility.

We measured the transmittance of the YBCO thin film at selected wavelengths — 450 nm, 500 nm, 530 nm, 580 nm, and 630 nm — using a mixed-polarized, variable-wavelength laser source. These measurements showed that the transmittance data at 630 nm matched the 633 nm measurements, as shown in figure 8, demonstrating repeatability and confirming the characteristic increase in transmittance during the superconducting transition. Across all measured wavelengths, the transmittance begins to rise near 100 K and levels off around 70 K, consistent with the behavior observed in figure 5b.

In addition to confirming the transition behavior, these measurements also revealed a wavelength dependence in the magnitude of the transmittance change. As listed in Table I, longer wavelengths exhibit more pronounced increases in transmittance, with up to 41 % relative change with respect to the room temperature transmittance as the sample transitions from the normal to the superconducting state. Even for shorter wavelengths, we observed a minimum relative change of 35 %.

| **Wavelength** (nm) | **Maximum Change in Transmittance** (%) **Relative to Room Temperature Transmittance** |
|---|---|
| 450 | $36 \pm 2$ |
| 500 | $35 \pm 2$ |
| 530 | $37 \pm 2$ |
| 580 | $41 \pm 2$ |
| 630 | $39 \pm 2$ |

Table I. Tabulated data for relative percentage change in transmittance of light from YBCO thin film for various wavelengths.

Similar to figure 7, the transmittance data at each wavelength were fitted with a logistic function to extract the midpoint of the sigmoid, corresponding to $T_c$. The resulting values are listed in Table

II, with $T_c$ estimates ranging from 80 K to 88 K across the different wavelengths. The uncertainty in $T_c$ at 450 nm is noticeably larger compared to other wavelengths, due to the smaller relative change in transmittance, which results in a less well-constrained logistic fit.

| Wavelength (nm) | Measured $T_c$ (K) |
|:---:|:---:|
| 450 | 84 ± 4 |
| 500 | 85 ± 1 |
| 530 | 84 ± 1 |
| 580 | 83 ± 1 |
| 630 | 82 ± 1 |

Table II. Tabulated data of mid-point values of critical transition temperature of YBCO thin film from the fit to a Logistic curve on transmittance data for various wavelengths of light source.

## IV. DISICUSSION

The electromagnetic response of a superconductor can be described phenomenologically by the Gorter and Casimir two-fluid model, which treats the electrical and optical properties defining charge carriers as a temperature dependent mixture of normal and superconducting fluids. Above the $T_c$, the electrical and optical properties are governed by the normal fluid and can be described by the Drude model. Below $T_c$, however, some fraction of the normal fluid converts into dissipation-free condensate that starts influencing the optical and electrical conductance [16]. Most of the previous investigations of the optical properties of YBCO and subsequent empirical models [17, 18] have concentrated in the far infrared or microwave region as these photon energies are closer to the superconducting gap. In these energy ranges the permittivity can come close to zero providing the possibility of interesting applications [19] in metamaterial or heterostructure sensors. Recently, an approach based on two-fluid model was used to explain enhanced transmission in the millimeter wavelength regime through an array of sub-wavelength holes in YBCO [20]. In the visible region, however, the condensate current is not expected to be dominant as the photon energies are much higher than the superconducting gap. However, our results show that the normal incidence reflectance and transmittance show a significant change near the superconducting transition temperature even in the visible region. We speculate that the formation of the superconducting fluid reduces absorption and, consequently, enhances the transmittance, explaining the transition seen in the transmittance data near $T_c$.

To qualitatively model this behavior, we employ an empirical formulation originally established for the microwave-frequency response of YBCO [17]. However, considering that optical frequencies are significantly higher than microwave frequencies, several approximations that are valid only in the microwave regime, particularly those assuming that the electromagnetic field oscillation period substantially exceeds the normal carrier relaxation time, are no longer appropriate. Therefore, we have removed these microwave-specific approximations and introduced additional terms into the model to ensure consistency with our optical-frequency experimental data.

The derived real ($\epsilon'_{\text{eff,relative}}$) and imaginary ($\epsilon''_{\text{eff,relative}}$) part of the dielectric permittivity based on our model, in terms of the normal fluid densities ($n_n$) and superconducting fluid densities ($n_s = n_0 - n_n$) are:

$$\epsilon'_{\text{eff,relative}} = \epsilon'_{\text{eff}}(\omega)/\epsilon_0 = \frac{\epsilon}{\epsilon_0} - \frac{n_0 - n_n(T,\hbar\omega)}{\mu_0 \epsilon_0 \lambda_L(0)^2 n_0 \beta \omega^2} - \frac{\frac{e^2 n_n(T,\hbar\omega)}{m_n}}{(\omega^4 + (\omega/\tau)^2)\epsilon_0}\omega^2 \quad (1)$$

$$\epsilon''_{\text{eff,relative}} = \epsilon''_{\text{eff}}(\omega)/\epsilon_0 = \frac{\frac{e^2 n_n(T,\hbar\omega)}{m_n}}{(\omega^4+(\omega/\tau)^2)\epsilon_0} \frac{\omega}{\tau}, \tag{2}$$

where $n_0$ is total fluid density, $T$ is the temperature, $\omega$ is the frequency of the incident light, $m_n$ is the effective mass of the normal fluid, $\lambda_L(0)$ is the London penetration depth at T = 0 K, $\tau$ is the relaxation time of the normal carriers, and $\beta$ is an introduced model parameter described below. $\lambda_L(0)$ is related to $\lambda_L(T)$, the London penetration depth at $T$ K, through the relation:

$$\left[\frac{\lambda_L(0)}{\lambda_L(T)}\right]^2 = 1 - \left(\frac{T}{T_c}\right)^\gamma, \tag{3}$$

Here $\gamma$ is a material-dependent exponent which critically depends on the nature of the superconductor under consideration. For conventional low-$T_c$ superconductors, $\gamma$ is usually equal to four [17]. However, for YBCO, values ranging between 1.2 - 2.5 have been found to represent the conductivity plots. However, this is critically dependent on how the YBCO was grown and in general shows a trend towards four with improved quality films.

The temperature-dependent relaxation time $\tau(T)$ is defined as:

$$\frac{1}{\tau(T)} = \begin{cases} \frac{1}{\tau(T_c)} \cdot \frac{T}{T_c}, & T \geq T_c, \\ \frac{1}{\tau(T_c)} \cdot \frac{\frac{T}{T_c}}{1+\alpha\left(\left(\frac{T}{T_c}\right)^{1-\gamma} - \frac{T}{T_c}\right)}, & T < T_c, \end{cases} \tag{4}$$

where $\tau(T_c)$ is the relaxation time at $T_c$, and $\alpha$ is an empirical parameter that points to any residual resistivity due to material imperfections.

As noted in Table I and Figure 9, the change in transmittance observed at $T_c$ is wavelength dependent. Specifically, at 450 nm the difference between the measured transmittance values at 12 K and 300 K is approximately half that observed for 630 nm. We have empirically introduced this behavior into the model by hypothesizing that a small fraction of the incident photon energy couples to superconducting carriers, converting them back to normal carriers.

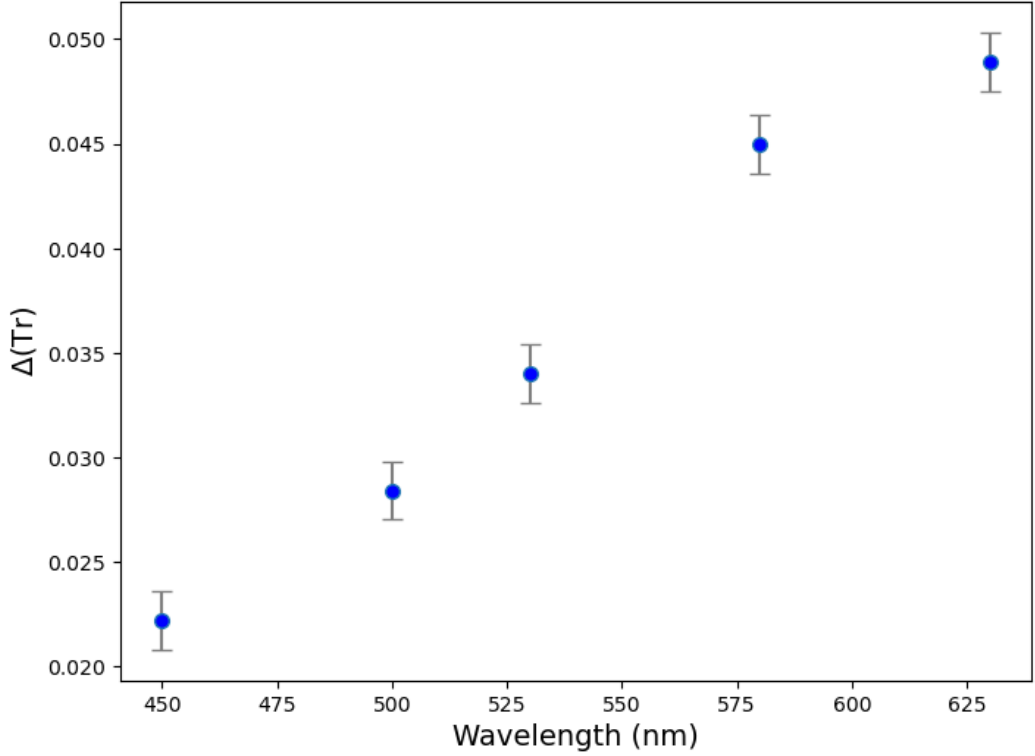

Figure 9. Change in transmittance (Δ(Tr)) across the superconducting transition as a function of wavelength. Δ(Tr) is the difference between the transmittance measured at 12 K and that measured at 300 K.

To take into account the conversion of the superconducting fluid to a normal fluid, we add an additional frequency dependence ($\beta(\hbar\omega)$) to the superconducting fluid density as,

$$n_s(T, \hbar\omega) = n_0 \cdot \left(1 - \left(\frac{T}{T_c}\right)^\gamma\right) \cdot \beta(\hbar\omega), \quad n_n(T, \hbar\omega) = n_0 - n_s(T, \hbar\omega). \tag{5}$$

where:

$$\beta(\hbar\omega) = 1 - e^{-\frac{E_{\text{gap}}}{\kappa\hbar\omega}}. \tag{6}$$

Here $\kappa$ represents the efficiency with which the photon energy is coupled to the superconducting fluid and $E_{gap}$ is a term reflective of the superconducting gap. This effect can be implemented in equation (1) and (2) by writing them entirely in terms of $n_n(T,\hbar\omega)$, which has been modified as:

$$n_n(T) = n_0 \begin{cases} 1, & T \geq T_c, \\ \left(1 - \beta + \beta\frac{T}{T_c}^\gamma\right), & T < T_c. \end{cases} \quad (7)$$

Using the $\epsilon'_{eff,\text{relative}(\omega)}$ and $\epsilon''_{eff,\text{relative}(\omega)}$ we can calculate the $n(\omega)$ and $k(\omega)$, which are the frequency dependent refractive indices and thus, obtain the temperature dependent transmittance for various wavelengths using the multi-layer transmittance equations given in [21].

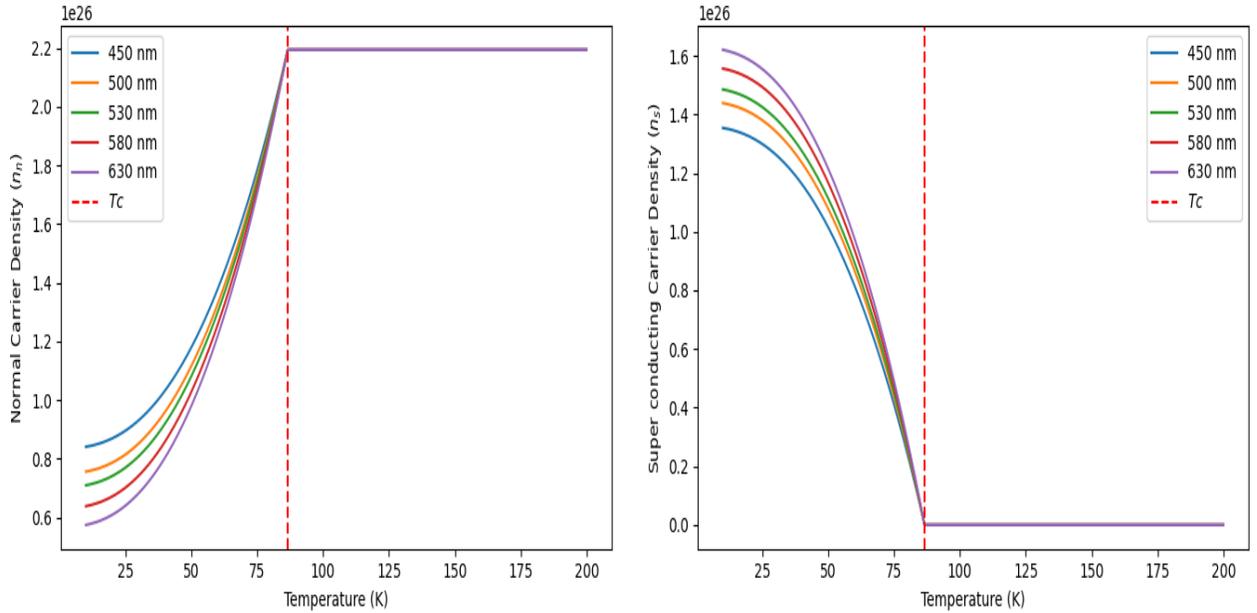

Figure 10. Variation of normal and superconducting current density as a function of temperature for various wavelengths.

The normal and superconducting fluid densities obtained from our model is shown in figure 10. The model depicts the reduction in the normal fluid density and the increase in the superconducting fluid density below $T_c$. As per the model, the imaginary part of the dielectric constant and thus, the imaginary part of the refractive index is dependent on the normal fluid density, and hence, it can be expected that the absorption within the material decreases resulting in an increase in transmittance, as shown in figure 11. We point out that the model is too simplistic to capture the superconducting transition and thus, the optical transition in a high- $T_c$ superconductor quantitatively. Further work will be required to quantify the observed variation in transmittance and reflectance across $T_c$.

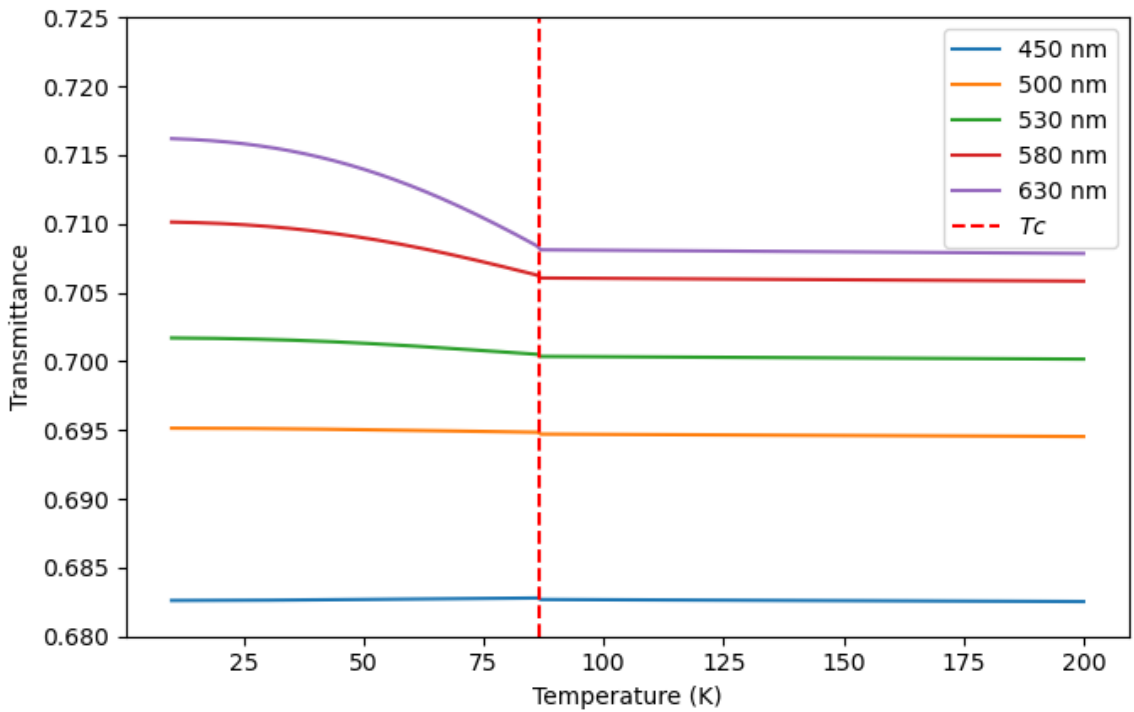

Figure 11. Transmittance as a function of temperature for various wavelengths simulated using the two-fluid model.

## V. CONCLUSION

In conclusion, we have investigated the optical response of YBCO thin films to visible light from 450 nm to 633 nm. By measuring the reflectance and transmittance as a function of sample temperature, across the superconducting phase transition, we have demonstrated a non-contact optical method for determining the $T_c$. This technique is particularly advantageous for small and thin film samples, where the establishment of electrical contacts for traditional resistivity measurements can be challenging. Detailed experimental results are presented for optical properties of YBCO thin films, showing distinct changes in reflectance and transmittance associated with the superconducting transition. Finally, an empirical model has been presented based on the two-fluid approach that qualitatively predicts the observed temperature-dependent transmittance and suggests that high-$T_c$ superconductor has potential applications in metamaterial or heterostructure sensors and as an optical readout scheme for superconducting devices.

**ACKNOWLEDGEMENT**

We would like to thank Dr. Muhammad N. Huda of UTA for insightful discussions on the optical properties of high-$T_c$ superconductors and the applicability of the two-fluid model. VAC acknowledge the support of NSF Grant No. CHE – 2204230.